\documentclass[aps,prb,twocolumn,superscriptaddress]{revtex4}
\usepackage{amsfonts}
\usepackage{amssymb}
\usepackage{amsmath}
\usepackage{graphicx}
\usepackage{epsfig}
\usepackage{color}

\begin{document}

\title{Dynamic Magnetization in Non-Hermitian Quantum Spin System}
\author{X. Z. Zhang}
\affiliation{College of Physics and Materials Science, Tianjin Normal University, Tianjin
300387, China}
\author{L. Jin}
\email{jinliang@nankai.edu.cn}
\author{Z. Song}
\email{songtc@nankai.edu.cn}
\affiliation{School of Physics, Nankai University, Tianjin 300071, China}

\begin{abstract}
We report a global effect induced by the local complex field, associated
with the spin-exchange interaction. High-order exceptional point up to ($N+1$)-level coalescence is created at the critical local complex field applied to the $N$-size quantum spin chain. The ($N+1$)-order coalescent level is
a saturated ferromagnetic ground state in the isotropic spin system.
Remarkably, the final state always approaches the ground state for an arbitrary
initial state with any number of spin flips; even if the initial state
is orthogonal to the ground state. Furthermore, the switch of macroscopic
magnetization is solely driven by the time and forms a hysteresis loop in
the time domain. The retentivity and coercivity of the hysteresis loop mainly rely on the non-Hermiticity. Our findings highlight the cooperation of non-Hermiticity and the interaction in quantum spin system, suggest a dynamical framework to
realize magnetization, and thus pave the way for the non-Hermitian quantum spin
system.
\end{abstract}

\maketitle



\section{Introduction}
Non-Hermitian quantum mechanics is an extension of
standard quantum mechanics and describes dissipative systems in a
minimalistic fashion \cite{Moiseyev2011}. The research field of non-Hermitian physics has been
greatly developed in the optical platforms \cite%
{Longhi2009,Feng2017,El-Ganainy2018,Midya2018,Miri2019,Oezdemir2019}; in particular, the
peculiar features of the exceptional point (EP) \cite{Miri2019}, which is the non-Hermitian phase transition point that solely presents in non-Hermitian system. The EP plays the
pivotal role in the intriguing dynamics and application including asymmetric mode switching \cite{Doppler2016},
unidirectional lasing \cite{Ramezani2014,Peng2016,Jin2018}, and enhanced optical
sensing \cite{Wiersig2014,Liu2016,Hodaei2017,Chen2017,Lau2018,Zhang2019,Lai2019,Hokmabadi2019}. Notably, the properties of EP highly depend on the level
coalescence and its topology \cite%
{Graefe2008,Ding2016,Xiao2019,Pan2019a}. Recently, the inspiring insights of
non-Hermitian physics emerge rapidly in the condensed-matter systems. The non-Hermitian quantum spin models \cite{Cejnar2007,Castro-Alvaredo2009,Giorgi2010,Zhang2013,Lee2014b,Ashida2017,Couvreur2017,Galda2018,Wang2020} and the exotic quantum many-body effect
ranging from non-Hermitian extensions of Kondo effect \cite%
{Nakagawa2018,Lourenfmmodeboxclsecio2018}, Fermi surface in coordinate space \cite{Mu2019}, Kibble Zurek mechanism \cite{Dora2019}, many-body localization \cite%
{Hamazaki2019}, to fermionic superfluidity \cite{Yamamoto2019,Okuma2019} are
reported. These findings unveil the interesting and important impacts of the
non-Hermiticity in the interacting systems.

In this paper, we uncover the influence of complex magnetic field in the
quantum spin system. Remarkably, we find that a local critical complex field
can induce the coalescence of substantial energy levels: the degenerate states with different symmetry of the
Hermitian quantum spin system coalesce at the critical complex field and form a high-order EP; the
order of coalescence is solely determined by the degeneracy. The
ground state associated with a saturated ferromagnetic ground state has
the highest order of coalescence and thus enables the dynamic magnetization.
For an initial state with any number of spin flips excited on the ground
state, the final state always approaches the ground state. Furthermore, a hysteresis loop is formed in the time domain and it is driven by the time rather than the magnetic field in contrast to the traditional magnetism. The properties of the non-Hermitian quantum spin system are capable of been examined from the retentivity and coercivity.

The rest of this paper is organized as follows: In Sec.~\ref{general} , we investigate the non-Hermitian quantum spin model, the non-Hermiticity of which stems from the complex magnetic field and propose a general method to connect the non-Hermitian model to the Hermitian spin model. With these preparations, in Sec.~\ref{local} we demonstrate that a local complex field can induce a global effect with the aid of the spin-exchange interaction. The formation of a high-order EP is therefore observed. Based on the performance of the dynamics of high-order EP, the dynamical generation of saturated ferromagnetic state and a hysteresis loop in the time domain are proposed in Sec.~\ref{dy1} and Sec.~\ref{dy2}. Sec.~\ref{conclusions} concludes this paper. Some details of
our calculation are placed in Appendix.

\section{Non-Hermitian quantum spin system}
\label{general}
In the quantum spin system, either a real or a complex field results in the splitting of the degenerate ground
states, where the spins are aligned along the direction of the external
field. However, the spectrum and the eigenstate of the system with a real
spectrum do not experience dramatic change in the present of the
external field; and the initial state exhibits a periodic oscillating
behavior among all the possible spin orientations. However, the situation
changes when a critical complex field is applied. The eigenstates coalesce
and the dynamics encounter dramatic changes in the sense that all the
initials state evolve to the coalescent state regardless of the initial spin
orientation. It is interesting to find out the intriguing features
of the quantum spin system in the presence of the complex field.

We consider a non-Hermitian spin system $H=H_{0}+H_{I}$ and show the
unique properties determined by the competition between the non-Hermiticity
and the interaction. The quantum spin system
\begin{equation}
H_{0}=-\sum_{i,j\neq i}(J_{ij}/2)\left(
s_{i}^{+}s_{j}^{-}+s_{i}^{-}s_{j}^{+}\right) +\sum_{i,j\neq i}\Delta
_{ij}s_{i}^{z}s_{j}^{z},  \label{n_H}
\end{equation}%
is subjected to an external complex field
\begin{equation}
H_{I}=\sum_{i}g_{i}\mathbf{h}\cdot \mathbf{s}_{i}.
\end{equation}%
The operators $s_{i}^{\pm }=s_{i}^{x}\pm is_{i}^{y}$ and $s_{i}^{z}$ are for
the spin-$1/2$ at the $i$-th site, obeying Lie algebra $%
[s_{i}^{z},s_{j}^{\pm }]=\pm s_{i}^{\pm }\delta _{ij}$ and $%
[s_{i}^{+},s_{j}^{-}]=2s_{i}^{z}\delta _{ij}$, where $\delta _{ij}$ is the
Dirac delta function. $\sum_{i,j\neq i}$ means the summation over all the
possible pair interactions at an arbitrary range. $J_{ij}$ represents the
inhomogeneous spin-spin interaction and $\Delta _{ij}$ characterizes the
anisotropy of the spin system $H_{0}$. The non-Hermiticity of $H_{I}$
originates from the complex magnetic field $\mathbf{h=}\left( 1,-i\gamma
,0\right) $, which can be understood as the spin-dependent losses and are
within the reach of ultracold atom experiments \cite%
{Lee2014,Ashida2017, Pan2019a}. The strength felt by each spin is $g_{i}\mathbf{h%
}$ in the inhomogeneous complex magnetic field. The system $H_{0}$ respects the
time-reversal symmetry $\mathcal{T}$ ($\mathcal{T}s_{i}^{\alpha }\mathcal{T}%
^{-1}=-s_{i}^{\alpha }$), which leads to the Kramers degeneracy when the
system possesses a half-integer total spin; the degeneracy breaks down
when the external complex field presents ($\mathcal{T}\mathbf{h}\cdot
\mathbf{s}_{i}\mathcal{T}^{-1}=-\mathbf{h}^{\ast }\cdot \mathbf{s}_{i}$).
The external field also spoils the commutation relation $%
[\sum_{i}s_{i}^{z},H]=0$.

\begin{figure}[tbp]
\centering
\includegraphics[width=0.49\textwidth]{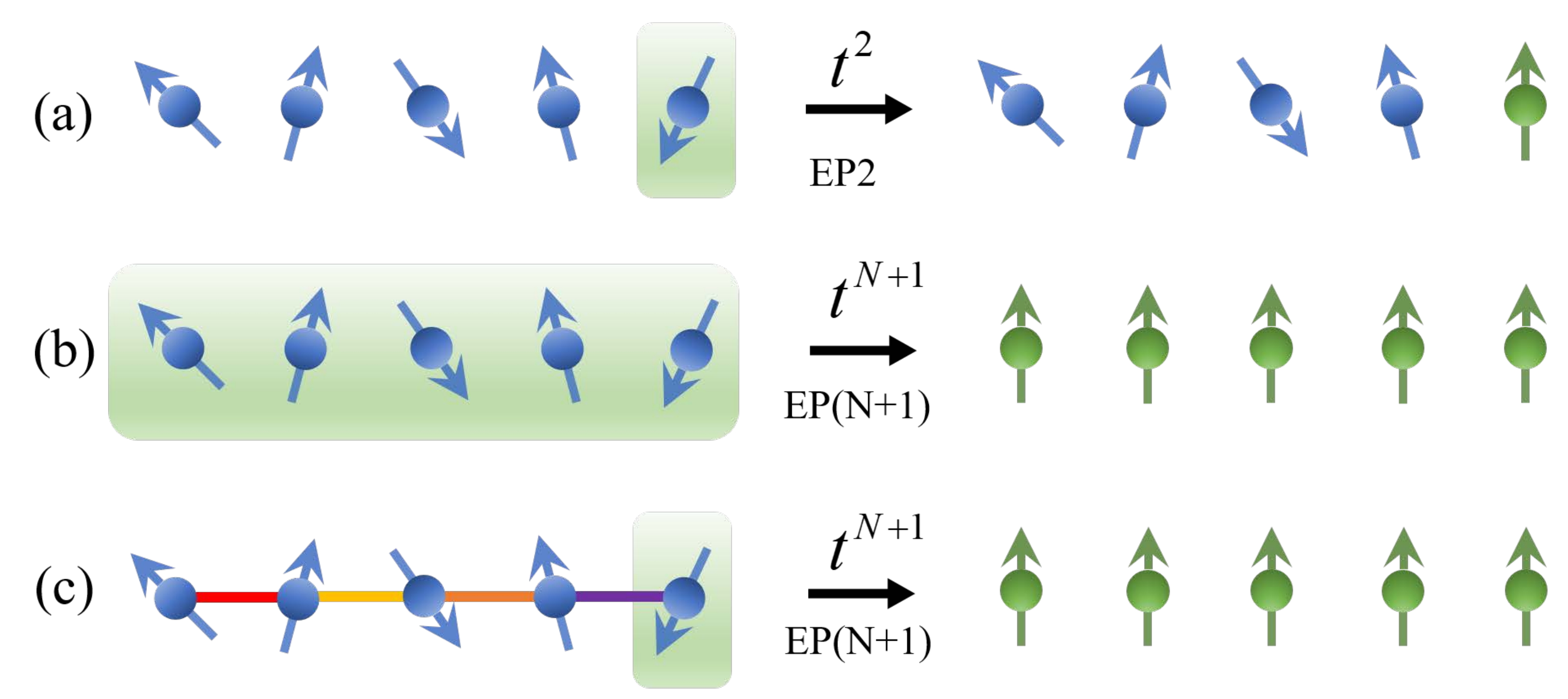}
\caption{Schematics of spins subjected to (a) a global complex field, (b) a
local complex field, and (c) a local complex field and interaction. The
complex magnetic field is shaded green. The couplings between different
spins are denoted by different colors representing inhomogeneous coupling $J_{ij}$%
. Two states coalesce in (b) and $N$ states coalesce in (a) and (c). Local
complex field only affects local spin without interaction, but can affect
globally with interaction.}
\label{fig_spin}
\end{figure}

We first show that a \textit{local} complex field dramatically changes the
ground state property of a quantum spin system. The Hilbert space of the non-Hermitian
system $H$ cannot be decomposed into subspaces in which the spin number is
specified even if $H_{0}$ has homogeneous spin-spin interaction $J_{ij}=J$, $%
\Delta _{ij}=\Delta $, and $\Delta \neq J$ (the XXZ model \cite{Yang1966}).
Considering a local complex field $H_{I}=g_{N}\mathbf{h}\cdot \mathbf{s}_{N}$
[Fig. \ref{fig_spin}(a)], $H_{0}$ and $H_{I}$ share two eigenstates even
though $\left[ H_{0},H_{I}\right] \neq 0$. The pair of eigenstates
\begin{equation}
\left\vert \psi \right\rangle _{xxz,\pm }=\pm \sqrt{1-\gamma }\left\vert
\Uparrow \right\rangle +\sqrt{1+\gamma }\left\vert \Downarrow \right\rangle ,
\end{equation}%
satisfy $H\left\vert \psi \right\rangle _{xxz,\pm }=(-N\Delta /4\pm \sqrt{%
1-\gamma ^{2}})\left\vert \psi \right\rangle _{xxz,\pm }$. Notably, $%
\left\vert \psi \right\rangle _{xxz,\pm }$ coalesce at $\left\vert \gamma
\right\vert =1$ (see Supplemental Material A). This indicates that the
ground state is dramatically changed from degeneracy to coalescence by the
local complex field.

Under a \textit{homogeneous global }complex field [Fig. \ref{fig_spin}(b)],
the Hamiltonian of the free spins in the absence of the interaction
describes a $\mathcal{PT}$-symmetric hypercube graph of $N$ dimension and
the system can be projected onto several invariant subspaces denoted by $s$ (%
$s=N/2$, $N/2-1$, $\cdots $) \cite{Zhang2012}. $\gamma =1$ is the EP$n$ ($%
n=2s+1$) of $n$-eigenstate coalescence in each subspace.
If the complex field is inhomogeneous, for example, the critical complex
field is locally applied to only a single spin, $\gamma =1$ reduces to an
EP2 of two-state coalescence \cite{Dembowski2001,Heiss2012}. To gain more insights for the interacting
spins under the homogeneous global complex field, we invite the exactly
solvable non-Hermitian Ising model to show that all the energy levels and
the eigenstates can be significantly affected by the complex field \cite{Zhang2015,Li2014}. To
proceed, we introduce a similarity transformation $\mathcal{S}%
=\prod\nolimits_{j}\mathcal{S}_{j}$, where $\mathcal{S}_{j}=e^{-i\theta
s_{j}^{z}}$ represents a counter-clockwise spin rotation in the $s_{x}$-$%
s_{y}$ plane around the $s_{z}$-axis by an angle $\theta $. Here $\theta
=\tan ^{-1}\left( i\gamma \right) $ is a complex number depending on the
strength of the complex field. Notably, the spin-rotation $\mathcal{S}_{j}$
is valid at arbitrary $\gamma $ unless at the EP of $H_{I}$, where $\mathbf{h%
}\cdot \mathbf{s}_{i}$ is in a nondiagonalizable Jordan block form. Under
the spin-rotation, $H$ is transformed to $\bar{H}=H_{0}\left( s\rightarrow
\tau \right) +\sqrt{1-\gamma ^{2}}\sum_{i}g_{i}\tau _{i}^{x}$, where the new
set of operators $\tau _{j}^{\pm }=\mathcal{S}_{j}s_{j}^{\pm }\mathcal{S}%
_{j}^{-1}$ and $\tau _{j}^{z}=\mathcal{S}_{j}s_{j}^{z}\mathcal{S}_{j}^{-1}$
also satisfies the Lie algebra, that is, $[\tau _{i}^{z},\tau _{j}^{\pm }]=\pm
\tau _{i}^{\pm }\delta _{ij}$ and $[\tau _{i}^{+},\tau _{j}^{-}]=2\tau
_{i}^{z}\delta _{ij}$. Notice that $\tau _{j}^{\pm }\neq (\tau _{j}^{\mp
})^{\dagger }$ due to the complex rotation angle $\theta $. We set $\left\{
\left\vert \psi _{n}\right\rangle \right\} $ as the eigenstates of the
operator $\sum_{i}s_{i}^{z}$ that represents all the possible spin
configurations along the $+z$ direction. Under the biorthogonal basis of $\{%
\mathcal{S}_{j}^{-1}\left\vert \psi _{n}\right\rangle \}$ and $\{\mathcal{S}%
^{\dagger }\left\vert \psi _{n}\right\rangle \}$, the matrix form of $\bar{H}
$ is Hermitian for $\left\vert \gamma \right\vert <1$. This directly leads
to an entirely real spectrum of $\bar{H}$. It is worth pointing out that the
transformation depends on $\gamma $ only and hence the spectrum is entirely
real even though a non-zero $g_{i}$ presents. This indicates that the
presence of the local complex field breaks the SU(2) symmetry of the system
but remains the entirely real spectrum without symmetry protection. In
general case, the EP of $H$ or $\bar{H}$ may not be the EP of $H_{I}$ at $%
\left\vert \gamma \right\vert =1$; however, we prove that $\left\vert \gamma
\right\vert =1$ is the EP of the non-Hermitian Ising model $H$ by taken $%
J_{ij}=0$, $\Delta _{ij}=1$ and $g_{i}=g$ (see Supplemental Material B) and
all the eigenstates coalesce. In particular, there are two types of
phase transitions in the non-Hermitian Ising model, the $\mathcal{PT}$
symmetry breaking at $\gamma =1$ and the spontaneous symmetry breaking at $g%
\sqrt{1-\gamma ^{2}}=1$, are both modulated by the transverse complex field $%
g$. Therefore, the \textit{homogeneous global} complex field can induce the phase
transitions of the non-Hermitian spin systems.

The interplay between the complex field and the spin-spin interaction brings
intriguing change to the system properties. In the aforementioned cases, the
influences of the local and global complex fields are discussed,
respectively. It is counter-intuitive that the \textit{local} field
associated with the \textit{inhomogeneous} interaction can generate the
effect induced by the \textit{homogeneous global} complex field; for
example, the strongly coalesced high-order EP($N+1$) for an $N$-spin system
can present in the quantum spin systems under a local complex field.

\section{High-order EP under local complex field}
\label{local}
In the XXZ model ($\Delta>J$), the ground states are two degenerate ferromagnetic states with all the spins
aligned in the $+z$ and $-z$ directions; and two degenerate ground states
coalesce at the critical complex field as an EP2. The underlying mechanism
of cooperation between the local complex field and spin-spin interaction
is elaborated as follows. For the ground state of the quantum spin system ($%
H_{0}$), the spin-spin interaction drives all the spins to behave like one
spin. If a spin is subjected to a complex field, all the spins feel the same
complex field because of the spin-correlation. Thus, all the degenerate
ground states of the quantum spin system coalesce at the critical complex
field. This enlightens us to propose the high-order EP. For a ferromagnetic
Heisenberg chain of $N$ spins, the ground states are ($N+1 $)-fold
degenerate with the angular momentum $N/2$ such that the projection of the spin
has $N+1$ values in an arbitrary direction. The response of the Hermitian system to
the external field can be ascribed to the performance of such an angular
momentum formed by $N$ non-interacting spins under the global complex field.
Therefore, the critical complex field turns all the possible spin
orientations to the $+z$ direction and an EP($N+1 $) is formed [Fig. \ref{fig_spin}%
(c)].

We consider the local complex field $g_{j}\mathbf{h}$ applied to the spin $%
\mathbf{s}_{j}$ in an isotropic ferromagnetic Heisenberg model with $\Delta
_{ij}=J_{ij}$, which is also referred to as the XXX model \cite%
{Heisenberg1928}. The spin-$1/2$ Heisenberg antiferromagnet often serves as
an effective low-energy description of the half-filled Hubbard model with
interaction. $H_{0}$ is rotationally invariant since it commutes with all
the three components of the total spin $\mathbf{s}=\sum_{j=1}^{N}\mathbf{s}%
_{j}$. Thus, the eigenstates of $H_{0}$ can be classified in terms of the total
spin number $s$. For the ferromagnetic Heisenberg model $H_{0}$, a saturated
ferromagnetic state, denoted as $\left\vert \Downarrow \right\rangle $, is
the members of the ground state multiplet \cite{Heisenberg1928,Yang1966}.
All the other degenerate ground states can be obtained by acting $s_{j}^{+}$
on $\left\vert \Downarrow \right\rangle $ step by step; the ground state is $%
(N+1)$-folder degenerate.

The Hermitian $H_{0}$ commutes with the non-Hermitian $H_{I}$ for the
homogeneous global critical complex field; however, the local complex field is a
nontrivial case since $H_{I}$ does not commute with $H_{0}$. The local
complex field breaks not only the spin-rotation symmetry, but also the
time-reversal symmetry; thus, all the eigenstates including the ground state
become non-degenerate. In principle, $H_{I}$ and $H_{0}$ do not share common
eigenstates and we cannot infer the property of $H$ from $H_{I}$. However,
an EP$(N+1$) emerges when $\left\vert \gamma \right\vert \rightarrow 1$.
This is observed from the subspace spanned by $\left\{ \left\vert
G_{n}\right\rangle \right\} $ that belongs to the subspace $s=N/2$, where $%
\left\{ \left\vert G_{n}\right\rangle \right\} $ is given by $\left\vert
G_{n}\right\rangle =(\sum_{i}s_{i}^{-})^{n-1}\left\vert \Uparrow
\right\rangle $, ($n=1,$ $2,$ $...$ $N+1$). $\left\{ \left\vert
G_{n}\right\rangle \right\} $ are the degenerate groundstates of $H_{0}$ of $%
(N+1)$-folder degeneracy with all the spins aligned in the same direction.
The condition of $\left\vert \gamma \right\vert \rightarrow 1$ guarantees
the validity of the perturbation theory in the representation of $\bar{H}$
and yields the matrix form of $H_{I}$ in the form of $W_{m,n}=g_{j}\sqrt{%
\left( N+1-m\right) m}[\left( 1+\gamma \right) \delta _{m+1,n}+\left(
1-\gamma \right) \delta _{m,n+1}]/2N$ (see Supplemental Material C). It is a
non-Hermitian hypercube with an EP$(N+1)$ at $\left\vert \gamma \right\vert
=1$ \cite{Zhang2012}. Similar as the exactly solvable non-Hermitian Ising
model, $\left\vert \gamma \right\vert =1$ is also the EP of the
non-Hermitian ferromagnetic Heisenberg model $H$.
\begin{figure*}[tbp]
\centering
\includegraphics[width=0.9\textwidth]{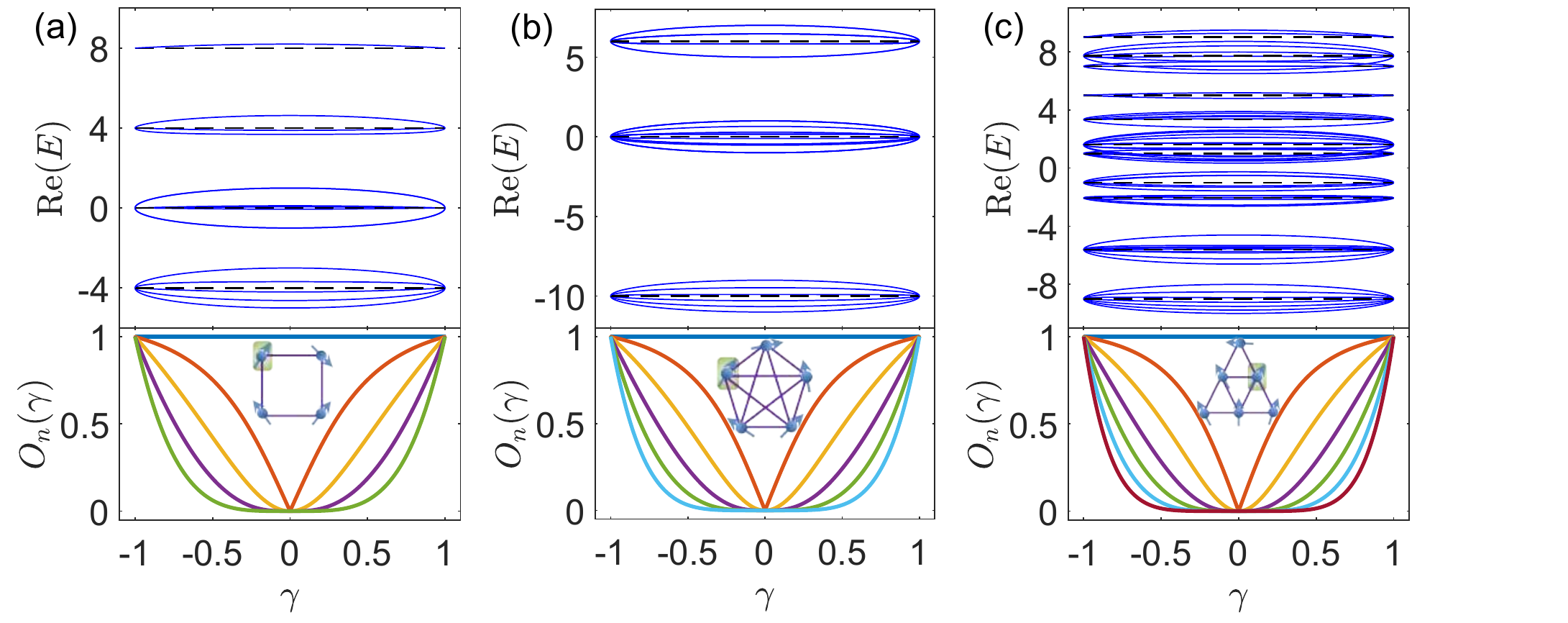}
\caption{Spectrum and degree of eigenstate similarity of three typical
non-Hermitian isotropic ferromagnetic Heisenberg models. (a) $N=4$, (b) $N=5$%
, (c) $N=6$. The insets are the schematics. The spin under the local complex
field is shaded green. The black dashed line denotes the eigen energy of $%
H_{0}$. The degeneracy of each black dashed line depends on the SU(2)
symmetry of the system.}
\label{fig_s}
\end{figure*}

We plot the eigenenergies of $H$ as functions of $\gamma $ in Fig. \ref%
{fig_s} to show the high-order EP. $O_{n}$ is introduced to quantify the
similarity between the excited state and the ground state; $O_{n}$ is
defined as
\begin{equation}
O_{n}=\left. \left\vert \langle \phi _{1}\left( \gamma \right) |\phi
_{n}\left( \gamma \right) \rangle \right\vert \right/ (\left\vert \phi
_{1}\left( \gamma \right) \right\vert \left\vert \phi _{n}\left( \gamma
\right) \right\vert ),
\end{equation}%
where $|\phi _{n}\left( \gamma \right) \rangle $ ($n=1,...N+1$) are the
ground state and $N$ excited states of $H$. The overlaps $O_{n}\left(
\left\vert \gamma \right\vert \right) \rightarrow 1$ in the plots. A
critical local magnetic field not only drives all the ground state
coalescence, but also the excited states coalesce at different energies with
multiple types of level coalescences and degeneracies, the order of level
coalesce is determined by the degenerate levels of $H_{0}$. Thus, the ground
state has the highest order of $(N+1)$-level coalescence. Furthermore, Figs. %
\ref{fig_s}(b) and \ref{fig_s}(c) clearly show that the complicated spin
structures induced by the interaction can harbour the high-order EPs
regardless of the position of applied external complex field. This is
a consequence that the ground states degeneracy is independent of the
spin-configuration \cite{Heisenberg1928,Yang1966}. This feature indicates our findings universally present.
\section{Dynamical generation of saturated ferromagnetic state}
\label{dy1}
The high-order EPs generated by the cooperation of the local complex
field and the spin-spin interaction brings intriguing dynamics. The spectrum
of the ferromagnetic Heisenberg model at the critical complex field is
constituted by many coalesced levels; instead of being diagonalized, the
system can be decomposed only into multiple Jordan blocks of various orders.
In each subspace, an arbitrary initial state will evolve towards the
coalescent state and its probability increases over time in power law according to the
order of coalescence. The more levels coalesced to one, the higher order of
the coalescence, and the faster probability increased in the dynamics. For an arbitrary initial state, the highest order of
the coalescent state determines the final state for a long time interval. Consequently, the
saturated ferromagnetic state is dynamically generated.

We consider a homogeneous spin-spin interaction $J_{ij}=J$ in the
ferromagnetic Heisenberg model. The coalesced ground state is $\left\vert
\Uparrow \right\rangle $ with all the spins aligned in the $+z$ direction.
The initial state finally approaches the saturated ferromagnetic state
because that the ground state has the highest order of coalescence at the
critical complex field and its probability increases dominantly in the
time-evolution process. For an arbitrary initial state $\sum_{n}c_{n}\left(
0\right) \left\vert G_{n}\right\rangle $ within the subspace $s=N/2$, the
coefficient $c_{m}\left( t\right) $ is
\begin{eqnarray}
c_{m}\left( t\right) &=&c_{m}\left( 0\right) +\sum_{n\neq m}\left(
-it/N\right) ^{n-m}\left( n-m\right) !h\left( n-m\right)  \notag \\
&&\times \lbrack \prod\limits_{p=m}^{m-1}p\left( N+1-p\right)
]^{1/2}c_{n}\left( 0\right) .
\end{eqnarray}%
where $h\left( n-m\right) $ is the Heaviside step function (see Supplemental
Material D). It indicates that the coefficient $c_{1}\left( t\right) $ of
the evolved state always includes the highest power of time $t$. Thus, the
component $c_{1}\left( t\right) $ of the evolved state overwhelms the other
components and the final state is the coalescent state $\left\vert \Uparrow
\right\rangle $. The fidelity $F\left( t\right) =$ $|\left. \left\langle
\Uparrow \right\vert e^{-iHt}\left\vert \Downarrow \right\rangle \right/
\left\langle \Uparrow \right\vert e^{-2s_{1}^{y}t}\left\vert \Downarrow
\right\rangle |^{2}$ $=[1+1/\eta ^{2}\left( t\right) ]^{-N}$ captures the
full dynamics, where $\eta \left( t\right) =t/t_{o}$ and $t_{o}=N/g_{1}$.
Obviously, $F\left( t\rightarrow \infty \right) =1$ and any inital state
with arbitrary spin flip evolves to $\left\vert \Uparrow \right\rangle $
(see Supplemental Material D). This feature is important for the hysteresis
loop in the time domain.

\section{Hysteresis loop in the time domain}
\label{dy2}
We consider a time-reversal process of two types of the $s^{z}=0$ initial states and observe their dynamics. The first type of initial state is a Neel state $\left\vert \Psi
_{\mathrm{I}}\left( 0\right) \right\rangle =\left\vert \uparrow \downarrow
\uparrow \downarrow \cdots \uparrow \downarrow \right\rangle $ and the
second type of initial state $\left\vert \Psi _{\mathrm{II}}\left( 0\right)
\right\rangle $ is the ground state of isotropic Heisenberg model with $%
s=N/2 $. We inspect the time dependent
average magnetization
\begin{equation}
M_{\mathrm{I}\left( \mathrm{II}\right) }\left( t\right) =N^{-1}\left.
\langle \Psi _{\mathrm{I}\left( \mathrm{II}\right) }\left( t\right) |\sigma
^{z}|\Psi _{\mathrm{I}\left( \mathrm{II}\right) }\left( t\right) \rangle
\right/ |\Psi _{\mathrm{I}\left( \mathrm{II}\right) }\left( t\right) |,
\end{equation}%
where $\left\vert \Psi _{\mathrm{I}\left( \mathrm{II}\right) }\left(
t\right) \right\rangle $ is the time-evolving state driven by the
non-Hermitian Heisenberg Hamiltonian.
\begin{figure*}[tbp]
\includegraphics[width=0.9\textwidth]{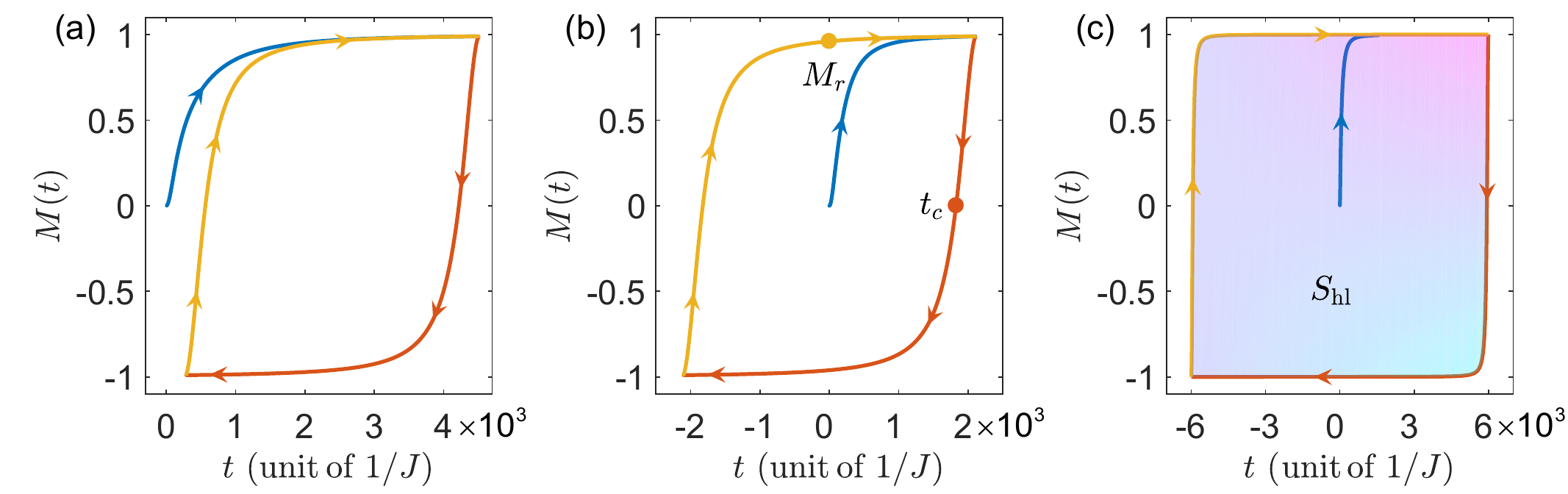}
\caption{Hysteresis loops for the initial state $\left\vert \Psi _{\mathrm{I}%
}\left( 0\right) \right\rangle $ in (a) and $\left\vert \Psi _{\mathrm{II}%
}\left( 0\right) \right\rangle $ in (b)-(c). The critical local complex
field $g_{1}$ is 0.02 in (a) and (b), and 0.1 in (c). The relaxation time is
$t_{f}=2\times10^3 J^{-1}$ in (a) and (b), and $t_{f}=3\times10^3 J^{-1}$ in
(c).}
\label{fighl}
\end{figure*}

The local complex field applied is in the form of $\mathbf{h=}\left(
1,i,0\right) $. We notice that any initial state finally evolves to the
saturated ferromagnetic state after a specified relaxation time $t_{f}\gg
t_{o}$. In Fig. \ref{fighl}, the dynamic magnetization is presented. The
trajectory of the initial state that has never been previously magnetized
follows the blue curve. After a relaxation time $t_{f}$, all the spins are
aligned along the $+z$ direction; keep increasing $t$ will produce slight
increase in $M(t)$. Then, we take the time-reversal action on $H$ and
observe the inverse magnetization. The trajectories are obtained as (see
Supplemental Material D)
\begin{equation}
M_{\pm }\left( t\right) =\pm \lbrack 1-\eta ^{2}\left( t\mp t_{f}\right)
]/[1+\eta ^{2}\left( t\mp t_{f}\right) ],
\end{equation}%
for the magnetization $M_{-}\left( t\right) $ (in yellow) and the inverse
magnetization $M_{+}\left( t\right) $ (in red). The red curve corresponds to
the inverse of the yellow curve. They form the hysteresis loop in the time
domain and is independent of the initial state. The stark difference from
the traditional hysteresis loop is that the switch of macroscopic
magnetization is driven in the time domain rather than the external field.
When $t$ is reduced to zero, some magnetic flux remains in the evolved state
even though the time is back to the origin. Similar to the hysteresis loop
driven by the external field, a non-zero magnetization point can be dubbed
as retentivity $M_{\text{\textrm{r}}}=1-2/(1+t_{o}^{2}/t_{f}^{2})$ (red
solid circle) in Fig. \ref{fighl}(b) and indicates the remanence of residual
magnetism in the evolved state. As time goes on, the yellow curve approaches zero and the corresponding point is referred to as coercivity denoted by
yellow solid circle in Fig. \ref{fighl}(b). The time required to remove the
residual magnetism from the state is called the coercive time $t_{\text{%
\textrm{c}}}=t_{f}-t_{o}$. As $t_{f}\rightarrow \infty $, the hysteresis
loop becomes a rectangle [Fig. \ref{fighl}(c)] of width $2$ and length $%
2t_{f}$. The formation of the hysteresis loop in the non-Hermitian spin
model is based on the time-reversal breaking induced by local complex field.
The area enclosed in the hysteresis loop is $S_{\mathrm{hl}%
}=4[t_{f}-2t_{o}\tan ^{-1}\left( t_{f}/t_{o}\right) ]$. Evidently, $S_{%
\mathrm{hl}}$ depends on the selection of $t_{f}$ that is in stark
difference to the traditional hysteresis loop whose area is constant
associating with the existence of a reversible magnetization phase \cite%
{Bertotti1998}. Furthermore, $S_{\mathrm{hl}}$ is related to the energy dissipated
due to material internal friction that is associated with the irreversible
thermodynamic change. The larger the area is, the more energy losses. The
complex field that is often interpreted as an effective coupling of the
system to the environment. Therefore, the non-Hermiticity causes the so-called thermal effect in the interacting systems.

\section{Conclusions}
\label{conclusions}
We have demonstrated that the local complex magnetic
field can induce dramatical changes in the spectral and dynamics of the
quantum spin systems. The form of interaction significantly matters in
the cooperation between the non-Hermiticity and interaction. The
inhomogeneous spin-exchange interaction assisted a local complex field can affect
globally and homogeneously. Specifically, the ground state of the isotropic
Heisenberg model subjected to a critical local magnetic field is a high-order of coalescent ferromagnetic state that has the lowest geometric
multiplicity of \emph{one}; for any initial state with an arbitrary number of
spin flips, the final states always approaches the ground state. This
discovery opens an avenue for magnetizing the non-Hermitian quantum spin system. A hysteresis
loop is obtained in the time domain, the local complex field from the effective
coupling between the system and the environment is associated with the
irreversible thermodynamic change. This unique feature is insensitive to
both the interaction range and the initial state.

\acknowledgments We acknowledge the support of the National
Natural Science Foundation of China (Grants No. 11975166, No. 11975128, and
No. 11874225). X.Z.Z. is also supported by the Program for Innovative
Research in University of Tianjin (Grant No. TD13-5077).

\appendix
\label{appendix}

\section{Non-Hermitian 1D XXZ model}
We consider a 1D non-Hermitian XXZ model with nearest neighbour homogeneous
spin-spin interaction, the corresponding Hamiltonian can be obtained by
taking $J_{ij}=J$ and $\Delta _{ij}=J$
\begin{equation}
H_{0}=-\sum_{j}\frac{J}{2}\left(
s_{j}^{+}s_{j+1}^{-}+s_{j}^{-}s_{j+1}^{+}\right) +\sum_{j}\Delta
s_{j}^{z}s_{j+1}^{z}.
\end{equation}%
We confine our discussion to $J>0$ without loss of generality. The XXZ chain
is in the ferromagnetic Ising phase when $\Delta >J$ \cite{Yang1966}: the
ground state is the saturated state with all spins aligned in either the $+z$
or $-z $ direction, i.e., the classical ground state with magnetization $%
s^{z}=\pm N/2$, where $N$ is the number of sites. We denote the two
degenerate ground states by $\left\vert \Uparrow \right\rangle $ and $%
\left\vert \Downarrow \right\rangle $, respectively; in this phase, the spin
reflection symmetry $s_{j}^{z}\rightarrow -s_{j}^{z}$ of the XXZ model
breaks. When the external magnetic field is switched on, the superposition
of the two ground states in the direction of the complex field $\left\vert
\psi \right\rangle _{\mathrm{xxz,}\pm }$ are the unnormalized eigenstates of
$H_{I}$,%
\begin{equation}
\left\vert \psi \right\rangle _{xxz,\pm }=\pm \sqrt{1-\gamma }%
\left\vert \Uparrow \right\rangle +\sqrt{1+\gamma }\left\vert \Downarrow
\right\rangle .
\end{equation}%
We can check that
\begin{equation}
\left( H_{0}+H_{I}\right) \left\vert \psi \right\rangle _{xxz,\pm
}=\left( -\frac{N\Delta }{4}\pm \sqrt{1-\gamma ^{2}}\right) \left\vert \psi
\right\rangle _{xxz,\pm }.
\end{equation}%
Therefore the states $\left\vert \psi \right\rangle _{\mathrm{xxz,}\pm }$
are the common eigenstates of both $H_{0}$ and $H_{I}$. It should be noted
that when $\left\vert \gamma \right\vert =1$, two such common states
coalesce to either $\left\vert \Uparrow \right\rangle $ or $\left\vert
\Downarrow \right\rangle $. This indicates that $\left\vert \gamma
\right\vert =1$ is the EP for $H_{I}$ and $H$.

\section{Non-Hermitian 1D Ising model}

The non-Hermitian 1D Ising model is
\begin{equation}
H=\sum_{j=1}^{N}s_{j}^{z}s_{j+1}^{z}+g\left( s_{j}^{x}+i\gamma
s_{j}^{y}\right) .  \label{H_ising}
\end{equation}%
The periodic boundary condition $s_{j}^{x,y,z}=s_{j+N}^{x,y,z}$ is assumed.
Using the similar transformation, the Hamiltonian (\ref{H_ising}) can be
transformed to
\begin{equation}
\overline{H}=\sum_{j}\tau _{j}^{z}\tau _{j+1}^{z}+g\sqrt{1-\gamma ^{2}}\tau
_{j}^{x},
\end{equation}%
which is a standard Ising model with modulated tranverse field $g\sqrt{%
1-\gamma ^{2}}$. Note that such the transformation holds if and only if $%
\left\vert \gamma \right\vert \neq 1$.

To obtain $\overline{H}$, we first perform the Jordan-Wigner transformation
\begin{eqnarray}
\tau _{j}^{x} &=&\frac{1}{2}-\overline{d}_{j}d_{j}, \\
\tau _{j}^{y} &=&\frac{i}{2}\sum_{j<l}\left( 1-2\overline{d}_{j}d_{j}\right)
\left( \overline{d}_{j}-d_{j}\right) , \\
\tau _{j}^{z} &=&-\frac{1}{2}\sum_{j<l}\left( 1-2\overline{d}%
_{j}d_{j}\right) \left( \overline{d}_{j}+d_{j}\right),
\end{eqnarray}%
to replace the quasi spin operators by the new non-Hermitian operators $%
\overline{d}_{j}$ and $d_{j}$, where $\overline{d}_{j}=\mathcal{S}%
_{j}c_{j}^{\dagger }\mathcal{S}_{j}^{-1}$ $(d_{j}=\mathcal{S}_{j}c_{j}%
\mathcal{S}_{j}^{-1})$ and $c_{j}^{\dagger }$ $(c_{j})$ represents the
creation (annihilation) operator of the spinless fermion. The new operators
satisfy the fermionic anticommutation relation
\begin{equation}
\left[ \overline{d}_{j},d_{j^{\prime }}\right] _{+}=\delta _{jj^{\prime }}.
\end{equation}%
We note that the parity of the number of such fermions is a conservative
quantity such that the Hamiltonian can be expressed as%
\begin{equation}
\overline{H}=\left(
\begin{array}{cc}
\overline{H}_{+} & 0 \\
0 & \overline{H}_{-}%
\end{array}%
\right) ,
\end{equation}%
where
\begin{equation}
\overline{H}_{+}=\overline{H}_{-}-2\left( \overline{d}_{N}\overline{d}_{1}+%
\overline{d}_{N}d_{1}+\overline{d}_{1}d_{N}+d_{1}d_{N}\right) ,
\end{equation}%
and
\begin{eqnarray}
\overline{H}_{-} &=&\frac{1}{4}\sum_{j=1}^{N}\left[ 2g\sqrt{1-\gamma ^{2}}%
\left( 1-2\overline{d}_{j}d_{j}\right) \right.  \notag \\
&&\left. +\left( \overline{d}_{j}\overline{d}_{j+1}+\overline{d}_{j}d_{j+1}+%
\overline{d}_{j+1}d_{j}+d_{j+1}d_{j}\right) \right] ,
\end{eqnarray}%
are the corresponding reduced Hamiltonians in the invariant subspaces with
even and odd parity. $\overline{H}_{+}$ represents a fermionic ring threaded
by a half of the flux quantum. The single-particle energy in two subspaces
can be obtained by the same procedures and will have the same value when the
system approaches the thermodynamic limit. In the following, we only focus
on the even parity subspace. Taking the Fourier transformation
\begin{equation}
d_{j}=\frac{1}{\sqrt{N}}\sum_{k}d_{k}e^{ikj},\text{ }\overline{d}_{j}=\frac{1%
}{\sqrt{N}}\sum_{k}\overline{d}_{k}e^{-ikj},
\end{equation}%
where $k=2\pi \left( m+1/2\right) /N$, $m=0$, $1$, $2$, ..., $N-1$. In the Nambu representation, the Hamiltonian can be written as a compact
form%
\begin{equation}
\overline{H}_{+}=\sum_{0<k<\pi }\overline{\eta }_{k}\overline{H}_{+}^{k}\eta
_{k},
\end{equation}%
with $\overline{\eta }_{k}=(%
\begin{array}{cc}
\overline{d}_{k} & d_{-k}%
\end{array}%
)$, $\eta _{k}=(%
\begin{array}{cc}
d_{k} & \overline{d}_{-k}%
\end{array}%
)^{T}$ and
\begin{equation}
\overline{H}_{+}^{k}=\frac{1}{2}\left(
\begin{array}{cc}
\left( \cos k-\lambda \right) & i\sin k \\
-i\sin k & -\left( \cos k-\lambda \right)%
\end{array}%
\right) ,
\end{equation}%
where $\lambda =2g\sqrt{1-\gamma ^{2}}$ and the Hamiltionian $\overline{H}%
_{+}^{k}$ satisfies the commutation relation $[\overline{H}_{+}^{k},$ $%
\overline{H}_{+}^{k^{\prime }}]=0$ ensuring the $\overline{H}_{+}$ can be
diagonalized within each $k$ subspace. To this end, we introduce the
non-Hermitian Bogoliubov transformation
\begin{eqnarray}
\overline{b}_{k} &=&\cos \frac{\beta _{k}}{2}\overline{d}_{k}+i\sin \frac{%
\beta _{k}}{2}d_{-k}, \\
b_{k} &=&\cos \frac{\beta _{k}}{2}d_{k}-i\sin \frac{\beta _{k}}{2}\overline{d%
}_{-k},
\end{eqnarray}%
where $\beta _{k}=\tan ^{-1}[\sin k/\left( \lambda -\cos k\right) ]$ and $%
\overline{b}_{k}$ is the quasi fermionic operator obeying the
anticommutation relation $\left[ \overline{b}_{k},b_{k^{\prime }}\right]
_{+}=\delta _{kk^{\prime }}$. It results in the diagonal form of the Hamiltonian
\begin{equation}
\overline{H}_{+}=\sum_{k}\varepsilon _{k}\left( \overline{b}_{k}b_{k}-\frac{1%
}{2}\right) ,
\end{equation}%
with the single-particle eigen energy $\varepsilon _{k}=\sqrt{\lambda
^{2}+1-2\lambda \cos k}/2$. Evidently, it is a non-interacting Hamiltonian
and hence the corresponding spectrum is fully determined by the
single-particle energy. If $\left\vert \gamma \right\vert <1$, then the
single-particle energy is real. Correspondingly, the system respects complex
single-particle spectrum regardless of $k$ when $\left\vert \gamma
\right\vert >1$. On the other hand, the level repulsion $\lim_{\left\vert
\gamma \right\vert \rightarrow 1}\left( \partial \varepsilon _{k}/\partial
\gamma \right) =\infty $ is observed as $\left\vert \gamma \right\vert $
approaches $1$, which is a typical feature of EP. In this sense, $%
\left\vert \gamma \right\vert =1$ is the EP of both $H_{I}$ and $H$.

\section{Non-Hermitian 1D isotropic Heisenberg model}

The Hamiltonian of the non-Hermitian Heisenberg model under the external
field is%
\begin{eqnarray}
H_{0} &=&-\frac{1}{2}\sum_{i,j\neq i}J_{ij}\left(
s_{i}^{+}s_{j}^{-}+s_{i}^{-}s_{j}^{+}+2s_{i}^{z}s_{j}^{z}\right) , \\
H_{I} &=&\sum_{\left\{ i\right\} }g_{i}\mathbf{h}\cdot \mathbf{s}_{i},
\end{eqnarray}%
where $\left\{ i\right\} $ represents $n\in \left[ 1,N\right] $ random
numbers denoting that $n$ spins are subjected to the local complex fields,
respectively. The presence of inhomogeneous magnetic fields breaks the $SU(2)$
symmetry, that is $\left[ s^{\pm },H\right] \neq 0$. However, the two
Hamiltonians $H_{0}$ and $H_{I}$ commute with each other when the index $i$
runs over all the spins and the critical local fields are applied so that $%
H_{I}$ can be treated as either $s^{+}$ or $s^{-}$. Although the two
Hamiltonians share the common eigenstates, the property of the ground states
is not clear since $s^{\pm }$ is non-Hermitian rather than Hermitian
operator, which cannot guarantee the validity of the perturbation theory. In
the following, we first use the transformation $\mathcal{S}$ to obtain a
Hermitian matrix $\overline{H}$ and then demonstrate that the existence of
the high-order EP neither depend on how many local fields are applied
nor on the spin configuration of $H$. Applying the spin-rotation $\mathcal{S}$,
the considered Hamiltonian can be transformed as
\begin{eqnarray}
\overline{H} &=&\overline{H}_{0}+\overline{H}_{I}, \\
\overline{H}_{0} &=&-\frac{1}{2}\sum_{\left\langle i,j\right\rangle
}J_{ij}\left( \tau _{i}^{+}\tau _{j}^{-}+\tau _{i}^{-}\tau _{j}^{+}+2\tau
_{i}^{z}\tau _{j}^{z}\right) , \\
\overline{H}_{I} &=&\sqrt{1-\gamma ^{2}}\sum_{\left\{ i\right\} }g_{i}\tau
_{i}^{x}.
\end{eqnarray}%
In the basis of $\tau ^{z}=\sum_{i}\tau _{i}^{z}$, the matrix form of $%
\overline{H}$ is Hermitian such that all the approximation method in quantum
mechanics can be applied. When $\left\vert \gamma \right\vert \rightarrow 1$%
, $\sqrt{1-\gamma ^{2}}$ is a small number indicating the weak coupling between
the spin and magnetic field. Therefore, $\overline{H}_{I}$ in the new frame
can be treated as weak perturbations. We focus on the effect of $\overline{H}%
_{I}$ on the ground state $\left\{ \left\vert
G_{n}\right\rangle \right\} $ of $\overline{H}_{0}$. $\overline{H}_{0}$ is a
standard isotropic Heisenberg model and hence the ground state is $(N+1)$ fold%
-degeneracy which can be expressed as%
\begin{equation}
\left\vert G_{n}^{\prime }\right\rangle =(\sum_{i}\tau
_{i}^{-})^{n-1}\left\vert \Uparrow \right\rangle ^{\prime }\text{ }\left(
n=1,\text{ }2\text{ }...\text{ }N+1\right) \text{, }
\end{equation}%
where
\begin{equation}
\left\vert \Uparrow \right\rangle ^{\prime }=\mathcal{S}\left\vert \Uparrow
\right\rangle \text{, and }\left\vert \Uparrow \right\rangle
=\prod\limits_{i=1}^{N}\left\vert \uparrow \right\rangle _{i}\text{.}
\end{equation}%
$\left\vert G_{n}^{\prime }\right\rangle $ is also the eigenstate of $\mathbf{%
\tau }^{2}=\sum_{i}\mathbf{\tau }_{i}^{2}$ with $\tau =N/2$. Notice that the
existence of the degenerate ground states are independent of
spin-configuration \cite{Heisenberg1928,Yang1966}. With the spirit of
degenerate perturbation theory, the eigenvalues up to the first order are
determined by the matrix form of $\overline{H}_{I}$ in the subspace spanned
by $\left\{ \left\vert G_{n}^{\prime }\right\rangle \right\} $. For
simplicity, the corresponding perturbed matrix is referred to as $W^{\prime
} $ whose elements are given as $W_{m,n}^{\prime }=\langle \overline{G}%
_{m}^{\prime }|\overline{H}_{I}|G_{n}^{\prime }\rangle $. $\{\langle
\overline{G}_{m}^{\prime }|\}$ are the biorthogonal left eigenvectors in the form of
\begin{equation}
\langle \overline{G}_{m}^{\prime }|=\left\langle \Uparrow \right\vert
U^{-1}(\sum_{i}\tau _{i}^{+})^{m-1}\text{ }\left( m=1,\text{ }2\text{ }...%
\text{ }N+1\right) .
\end{equation}%
Here we stress two points: (i) One can always safely throw away high-order
correction when $\left\vert \gamma \right\vert \rightarrow 1$ due to the
Hermiticity of matrix $W^{\prime }$. (ii) When homogenous magnetic filed is
applied, that is $\left[ \overline{H}_{I},\text{ }\overline{H}_{0}\right] =0$%
, $\overline{H}$ can be decomposed into block matrix in light of the
eigenvector of $\mathbf{\tau }^{2}$ and hence the eigenvalues of $W^{\prime
} $ are the energies of groundstate and $N$ excited states of $\overline{H}$
in the unbroken region. After straightforward algebras, one can readily
obtain the entry of matrix $\ W_{m,n}^{\prime }=\sqrt{1-\gamma ^{2}}%
\sum_{\left\{ i\right\} }g_{i}[\sqrt{\left( N+1-m\right) m}\left( \delta
_{m+1,n}+\delta _{m,n+1}\right) ]/2N$, where $1/N$ stems from the
translation symmetry of the groundstate $\{\left\vert G_{n}\right\rangle
^{\prime }\}$. Performing the inverse transformation $W=\mathcal{S}%
^{-1}W^{\prime }\mathcal{S}$ ($W_{m,n}=\left\langle G_{m}\right\vert
\overline{H}_{I}\left\vert G_{n}\right\rangle $ with $\left\vert
G_{n}\right\rangle =\mathcal{S}^{-1}\left\vert G_{n}^{\prime }\right\rangle $%
), the element of matrix $W$ can be given as $W_{m,n}=\sum_{\left\{
i\right\} }g_{i}\sqrt{\left( N+1-m\right) m}[\left( 1+\gamma \right) \delta
_{m+1,n}+\left( 1-\gamma \right) \delta _{m,n+1}]/2N$. It is a non-Hermitian
hypercube \cite{Zhang2012}, the EPN will exhibit when $\gamma =1$.
Therefore, a local magnetic will lead to a high-order of EP, the order of
which is determined by the degeneracy of ground state energy of $H_{0}$.

\section{Dynamics at the EPN}

\subsection{Generating the saturated ferromagnetic state}

\begin{figure}[tbp]
\centering
\includegraphics[width=0.35\textwidth]{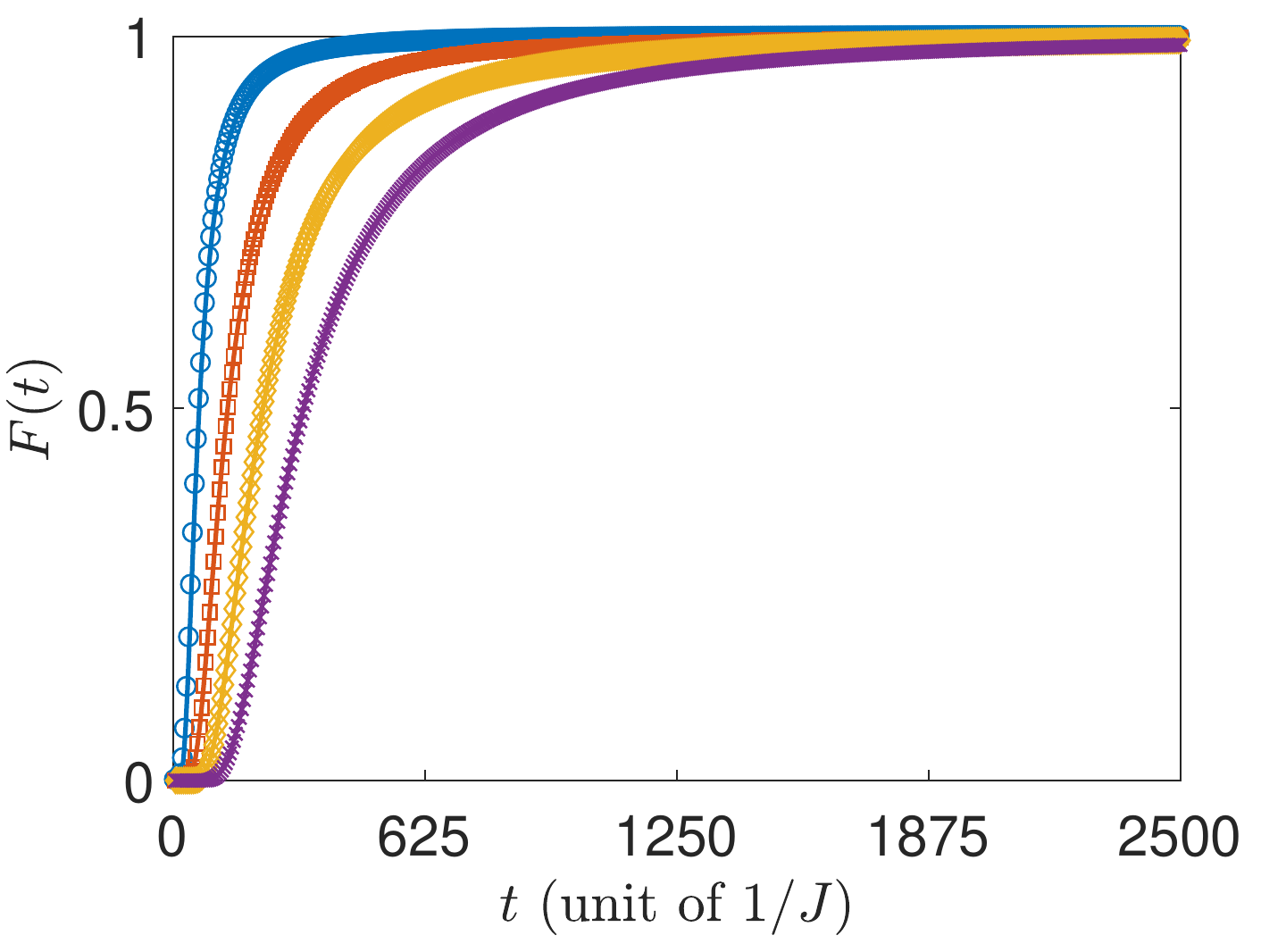}
\caption{Plots of the normalized fidelity $F\left( t\right) $ as functions
of time $t$ for the critical systems with $N=3$, $5$, $7$, and $9$ denoted
by the blue circle, red square, yellow diamond, and purple cross,
respectively. The circle denotes the corresponding analytical result
obtained by Eq. (\protect\ref{Ft}). $F\left( t\right) $ means the fidelity
driven by either $H$ (solid lines) or $W $ (hollow markers).}
\label{fig_trajectory}
\end{figure}

We show how to generate a saturated ferromagnetic state, where all local
spins (or conduction electron spins) are aligned parallel to the $z$-axis.
The non-Hermitian Heisenberg model Hamiltonian is given by Eqs. (1) with $%
\Delta _{ij}=J_{ij}$ in the main text. We assume the magnetic field is
applied to spin at site number $1$ unless stated otherwise, that is $i=1$.
When the single magnetic field is at the critical value $\gamma =1$, the
matrix form of $W$ can be given as $W_{m,n}=g_{1}\sqrt{\left( N+1-m\right) m}%
\delta _{m+1,n}/N$ that is a Jordan block form of dimension $N+1$. The
corresponding coalescent eigenstate is $\left\vert \Uparrow \right\rangle
=\prod\nolimits_{i=1}^{N}\left\vert \uparrow \right\rangle _{i}$. Notice
that $W$ is a nilpotent matrix with order $\left( N+1\right) $ such that $%
\left( W\right) ^{N+1}=0$. The element of matrix $W^{k}$ can be expressed as
\begin{equation}
\left( W^{k}\right) _{mn}=[\prod\limits_{p=m}^{m+k-1}p\left( N+1-p\right)
]^{1/2}\frac{g_{1}}{N}\delta _{m+k,n},  \label{ele_W}
\end{equation}%
where $k<N+1$. We focus on the dynamics of the critical system $W$. The
evolved state in this subspace is governed by the propagator $\mathcal{U=}%
e^{-iWt}$. With the aid of Eq. (\ref{ele_W}), one can readily obtain the
element of propagator
\begin{eqnarray}
\mathcal{U}_{m,n} &=&\delta _{mn}+\left( \frac{-itg_{1}}{N}\right) ^{n-m}%
\frac{h\left( n-m\right) }{\left( n-m\right) !}  \notag \\
&&\times \lbrack \prod\limits_{p=m}^{n-1}p\left( N+1-p\right) ]^{1/2},
\end{eqnarray}%
where $h\left( x\right) $ is a step function with the form of $h\left(
x\right) =1$ $\left( x>0\right) ,$ and $h\left( x\right) =0$ $\left(
x<0\right) $. Considering an arbitrary initial state $\sum_{n}c_{n}\left(
0\right) \left\vert G_{n}\right\rangle $, the coefficient $c_{m}\left(
t\right) $ of evolved state is
\begin{eqnarray}
c_{m}\left( t\right) &=&c_{m}\left( 0\right) +\sum_{n\neq m}\left( \frac{%
-itg_{1}}{N}\right) ^{n-m}\frac{h\left( n-m\right) }{\left( n-m\right) !}
\notag \\
&&\times \lbrack \prod\limits_{p=m}^{n-1}p\left( N+1-p\right)
]^{1/2}c_{n}\left( 0\right) .  \label{evolve_c}
\end{eqnarray}%
It clearly shows that no matter what the initial state is selected, the
coefficient $c_{1}\left( t\right) $ of evolved state always includes the
highest power of time $t$. As time goes on, the component $c_{1}\left(
t\right) $ of the evolved state overwhelms the other components ensuring the
final state is coalescent state $\left\vert \Uparrow \right\rangle $ under
the Dirac normalization. The different types of the initial state just
determine how the total Dirac probability of the evolved state increases
over time and the relaxation time for it evolves to the coalescent state. To
measure the similarity between evolved state and target ferromagnetic state $%
\left\vert \Uparrow \right\rangle $, we introduce the normalized fidelity as%
\begin{equation}
F_{n}\left( t\right) =\frac{\left\vert \left\langle G_{1}\right\vert
\mathcal{U}\left\vert G_{n}\right\rangle \right\vert ^{2}}{\left\langle
G_{n}\right\vert \mathcal{U}^{\dagger }\mathcal{U}\left\vert
G_{n}\right\rangle }.
\end{equation}%
The quantity $F_{n}\left( t\right) $ also reflects how fast the evolved
state approaches the final state. Using the Eq. (\ref{evolve_c}), one can
give directly the following expression
\begin{equation}
F_{n}\left( t\right) =\frac{\delta _{1,n}+\left( \eta ^{2}\right)
^{n-1}C_{N}^{n-1}}{1+\sum_{m=1}^{N}\left( \eta ^{2}\right)
^{n-m}C_{n-1}^{m-1}C_{N+1-m}^{n-m}},  \label{Ft}
\end{equation}%
where $\eta \left( t\right) =t/t_{o}$ with $t_{o}=N/g_{1}$. For $n=1$, the
initial state is the eigenstate of $W$ and hence does not evolve. We plot (%
\ref{fig_trajectory}) to compare the numerical result obtained by driving
the interacting Hamiltonian $H$ and analytical result based on the
perturbation matrix $W$. The initial state will evolve to the
target state and excellently agrees with our prediction. On the other hand,
as the system dimension increases, we observe that the initial state takes
more time to arrive at the final ferromagnetic state. Although the system
with larger sites can host higher-order EPs, the relaxation time is
inversely proportional to the order of EP that can be readily understood by
analytical formula $F_{N+1}\left( t\right) =[1+1/\eta ^{2}\left( t\right)
]^{-N}$. This again confirms the validity of the perturbation treatment of the
non-Hermitian external field.

\subsection{The hysteresis loop in time domain}

Another interesting dynamical phenomenon is the hysteresis loop in the time
domain. The hysteresis loop can be obtained through measuring the average
magnetization $M\left( t\right) $ [Eq. (6) in the main text]. Here we
demonstrate that two components of the loop (red and yellow line in Fig. 3
of the main text) can be derived analytically based on the aforementioned
mechanism. When the initial state is magnetized, the final state is
coalescent state in the subspace $W$ such that all the inverse magnetization
process is solely determined by the effective Hamiltonian $W$.
Correspondingly, $M\left( t\right) $ is given as%
\begin{equation}
M\left( \sigma ,t\right) =\frac{1}{N}\frac{\sum_{m}\mathcal{U}_{m,\sigma
}^{\ast }\mathcal{U}_{m,\sigma }\left( N-2m+2\right) }{\sum_{m}\mathcal{U}%
_{m,\sigma }^{\ast }\mathcal{U}_{m,\sigma }},
\end{equation}%
where $\sigma $ is either $N+1$ or $1$ and denotes the magnetization (yellow)
or inverse magnetization (red) process. Straightforward
algebras show that $M\left( N+1,t\right) =[1-\eta ^{2}\left( t\right)
]/[1+\eta ^{2}\left( t\right) ]$ and $M\left( 1,t\right) =[\eta ^{2}\left(
t\right) -1]/[1+\eta ^{2}\left( t\right) ]$. In the main text, Eqs. (7)-(8)
are obtained by replacing $\eta \left( t\right) $ with $\eta \left( t\mp
t_{f}\right) $ according to the starting point in the time-evolution process. Consequently, the physical
quantities coercive time $t_{\text{\textrm{c}}}$ and retentivity $M_{\text{%
\textrm{r}}}$ defined in main text can be given as%
\begin{equation}
t_{\text{\textrm{c}}}=t_{f}-t_{o},M_{\text{\textrm{r}}}=1-\frac{2}{%
1+t_{o}^{2}/t_{f}^{2}},
\end{equation}%
where $t_{f}$ is the relaxation time that the initial state being magnetized
(end point of the blue line). The area enclosed in the hysteresis loop is%
\begin{equation}
S_{\mathrm{hl}}=4[t_{f}-2t_{o}\tan ^{-1}\left( t_{f}/t_{o}\right) ].
\end{equation}%
We find that all the quantities of the hysteresis loop are associated with $%
t_{f}$. It is a unique feature of the considered non-Hermitian spin model
and is distinct from the traditional hysteresis loop. In the context of
magnetism, there exists a reversible magnetization phase in the whole
magnetization process so that the area of the hysteresis loop is independent of
relaxation time \cite{berto} . This suggests that no matter how one
increases the external field strength, the area surrounded by the loop is
always the same. However, the presence of local complex field spoils the
time-reversal symmetry and hence induce a hysteresis loop depending on $t_{f}
$. When $t_{f}/t_{o}\gg 1$, the hysteresis loop tends to a rectangle whose
width and length are $2$ and $2t_{f}$. Such a graph may inspire further
interest in the experiment.

\end{document}